\documentclass[twocolumn,preprintnumbers,amsmath,prl,amssymb,epsf]{revtex4}
\usepackage{graphicx}
\usepackage[dvips]{color}
\usepackage{newfloat}
\DeclareFloatingEnvironment[name={FIG.S}]{suppfigure}

\begin{document}
\title{Controllable stimulation of retinal rod cells using single photons}
\author{Nam Mai Phan$^{1,3}$, Mei Fun Cheng$^{1}$, Dmitri A. Bessarab$^{2}$, and Leonid A. Krivitsky$^{1,}$}
\email{Leonid_Krivitskiy@dsi.a-star.edu.sg}
\affiliation{$^1$ Data Storage Institute, Agency for Science Technology and Research (A-STAR), 117608 Singapore\\
$^2$ Institute of Medical Biology, Agency for Science Technology and Research (A-STAR), 138648 Singapore\\
$^3$ Department of Bioengineering, National University of Singapore, Singapore 117576} \vskip 24pt

\begin{abstract}
\begin{center}\parbox{14.5cm}
{New tools and approaches of quantum optics offer a unique opportunity to generate light pulses carrying a precise number of photons. Accurate control over the light pulses helps to improve the characterization of photo-induced processes. Here, we interface a specialized light source which provides flashes containing just one photon, with retinal rod cells of\textit{ Xenopus laevis} toads. We provide unambiguous proof of single photon sensitivity of rod cells without relying on the statistical modeling. We determined their quantum efficiencies without the use of any pre-calibrated detectors, and obtained the value of (29$\pm$4.7)\%. Our approach provides the path for future studies and applications of quantum properties of light in phototransduction, vision, and photosynthesis.}
\end{center}
\end{abstract}
%\pacs{42.50.Ar, 42.66.Lc,	87.80.Jg}
 \maketitle \narrowtext
\vspace{0mm}

\textcolor[rgb]{0,0,0}{Ability to control light at a quantum level can be extremely useful in addressing biological problems. Interfacing  biological objects with non-classical (quantum) light allows to enhance the precision of biological measurements \cite{Bahor}, fosters development of more precise models of biological processes \cite{Teich}, and allows to reveal possible role of quantum effects in neurobiology \cite{Sekatski, Pomarico} and perception \cite{Thaheld, Koch}.}

\textcolor[rgb]{0,0,0}{Rod cells of the retina are natural photodetectors, and they are perfect candidates for studies of biological interfaces with quantum light. Rod cells convert incident light into electrical currents, which are then sent to the brain via the optics nerve. They are responsive at discrete photon level and highly-sensitive techniques for readout of their electrical response are readily available \cite{Baylor,Bodoia}.}

To date, only classical light sources (lasers, lamps, light emitting diodes etc.) have been used in visual studies\cite{Baylorsingle,Review}. Quantum mechanics imposes a fundamental limit on the stability of such sources. The number of emitted photons is not fixed, but rather follows a defined probability distribution, depending on the light source \cite{Loudon}. \textcolor[rgb]{0,0,0}{This leaves doubt about the exact number of photons used to stimulate the rod cell.} Impact of \textcolor[rgb]{0,0,0}{unavoidable} photon fluctuations becomes a crucial issue for the experiments conducted at discrete photon level. Using the light source with a  "fixed" number of photons, would allow for more precise \textcolor[rgb]{0,0,0}{and direct} characterization of rod cells, and facilitate the development of more accurate mathematical models for vision and phototransduction processes.   

%Since the \textcolor[rgb]{1,0,0}{exact} number of photons (one or more) of each light pulse stimulating the rod cell is not known, statistical modeling is always required for interpretation of experimental results. A Poissonian model was used in earlier studies to suggest single photon sensitivity of rod cells \cite{Baylorsingle,Review}. However, the choice of the model may not be unique, and it should account for a large number of parameters which are specific for each given cell and inaccessible for direct verification \cite{Hamer,Caruso}. Suppression of photon fluctuations in the light source\textcolor[rgb]{1,0,0}{, which is possible using the tools and approaches of quantum optics,} will allow for more precise \textcolor[rgb]{1,0,0}{and direct} characterization of rod cells, and facilitate the development of more accurate mathematical models for vision and phototransduction processes.

A number of methods for reliable generation of light pulses with fixed numbers of photons \textcolor[rgb]{0,0,0}{(Fock states)} have been suggested \cite{Eisaman}. It was theoretically proposed to use such pulses for characterization of individual stages of the phototransduction \cite{Teich}, visual detection of quantum entanglement \cite{Sekatski,Pomarico}, and precise determination of the visual threshold \cite{Teich,Holmes}.

\textcolor[rgb]{0,0,0}{In this letter we experimentally realize a "noise-free" single photon light source and interface it with a biological object. Our experiment allows us to resolve several problems} which can not be addressed using classical light sources, including 1) demonstration of single photon sensitivity of rod cells without relying on statistical modeling, \textcolor[rgb]{0,0,0}{2) precise determination} of parameters of rod cells single photon responses without the interference from multiphoton detection events 3) \textcolor[rgb]{0,0,0}{accurate} measurement of the quantum efficiency of rod cells without pre-calibrated devices.

\textcolor[rgb]{0,0,0}{We exploit spontaneous parametric down conversion (SPDC) \cite{KlyshkoPhotons}, which is known to be one of the most accessible and versatile approaches to generation of single photons.} In the SPDC a photon of a laser pulse (pump), propagating in a nonlinear optical crystal, is converted with some probability ($\approx10^{-6}$) into a pair of photons (signal and idler), obeying conservation of energy and momentum: 
\begin{equation}
	\omega_{p}=\omega_{s}+\omega_{i};\\\\\\\	\vec{k}_{p}=\vec{k}_{s}+\vec{k}_{i}
\end{equation}
where $\omega_{p,s,i}$ and $\vec{k}_{p,s,i}$ are the frequencies and the wave vectors of pump, signal, and idler photons, respectively. 
Conservation laws (1) guarantee that signal and idler photons have well defined frequencies, and emission directions. In our experiment we use a Q-switched Nd:YAG laser (Crystalaser, $\lambda_p$=266 nm, pulse duration 30 ns, repetition rate 25 kHz) as a pump and a nonlinear 5 mm long $\beta$-barium borate (BBO) crystal. Signal and idler photons are emitted from the BBO in two directions, which form an angle of $\pm3^{\circ}$ to the direction of the pump, see Fig.1. They have the same wavelengths $\lambda_{s}$=$\lambda_{i}$=532 nm, which are chosen to maximize photon absorption by the rhodopsin photo-pigment in the cell \cite{Harosi,Palacios}.

Simultaneity in emission of signal and idler photons is used for generation of single photon pulses \cite{Rarity}. The signal photon is addressed to a single photon avalanche photodiode (APD, Perkin-Elmer). The APD output is used as a trigger for an acousto-optical modulator (AOM, Gooch and Housego) in the idler beam, see Fig.1 . Once the signal photon is detected by the APD, the AOM is activated for a period of 100 ns, during which it diverts the idler photon to an optical fiber pointing at the rod cell. An idler photon is optically delayed by a 45 m long fiber to compensate for incurring delays. Details of the experiment synchronization are shown in Fig.S1 of Supplemental Material \cite{Sup}. If the APD does not detect a signal photon, the AOM remains inactive, and no light pulse is sent to the rod cell. 

Ideally each photocount of the APD in the signal beam heralds a single photon in the idler beam, which is directed to the rod cell. However, inefficiencies of optical elements in the idler beam lead to losses of some of the idler photons. We measured that the probability of a heralded idler photon to reach the rod cell is about 22\%. The detailed analysis of optical losses is presented in Supplemental Material \cite{Sup}.

Single photon sources are conventionally characterized with the second order correlation function $g^{(2)}$ 
\begin{equation} 
g^{(2)}=1+\frac{VarN-\left\langle N \right\rangle}{\left\langle N \right\rangle^{2}},
\end{equation}
where $\left\langle N \right\rangle$, and $VarN$ is the mean and the variance of the number of photons, respectively 
\cite{KlyshkoFoundations}. For Poissonian light sources $g^{(2)}$=1, whilst for an ideal single photon source $g^{(2)}$=0. We measure $g^{(2)}$ of light in the idler beam in the independent experiment, using a 50/50 fiber beamsplitter (Thorlabs), and two gated APDs (Perkin-Elmer). APD signals are addressed to a coincidence circuit (CC) with a time window of 120 ns (Phillips Scientific). Then $g^{(2)}\propto{N_c/(N_1N_2)}$, where $N_1$, $N_2$ are the numbers of APD photocounts, and $N_c$ is the number of photocounts coincidences \cite{KlyshkoFoundations}. Our measurement yields $g^{(2)}=0.08\pm0.06$. Thus the probability of emission of more than one photon is about 12 times smaller, compared to the Poissonian light source with the same mean photon number. Details on characterization of the single photon source are described in Fig.S2 of Supplemental Material \cite{Sup}. The obtained value of $g^{(2)}$ compares favorably to the ones typically obtained with \textcolor[rgb]{0,0,0}{alternative} single photon sources \cite{Eisaman}.

\textcolor[rgb]{0,0,0}{Our light source also provides the possibility of measurement of quantum efficiency of rod cells \cite{Burnham,Malygin,Migdall}. The quantum efficiency $\eta$ characterizes the ability of rod cells to respond to the impinging light, and it is defined as: 
\begin{equation}
\eta\propto{R/N_{APD=1}},
\end{equation}
where $R$ is the number of rod cell responses, and $N_{APD=1}$ is the number of incident photons, which is proportional to the number of photocounts of the APD in the signal beam. In contrast to the conventional approach, it is a direct method of measurement, which does not require calibration of the photometer, optical standards, and it does not rely on the choice of any particular model of rod cell response.}

\begin{figure}
			\includegraphics[width=\linewidth]{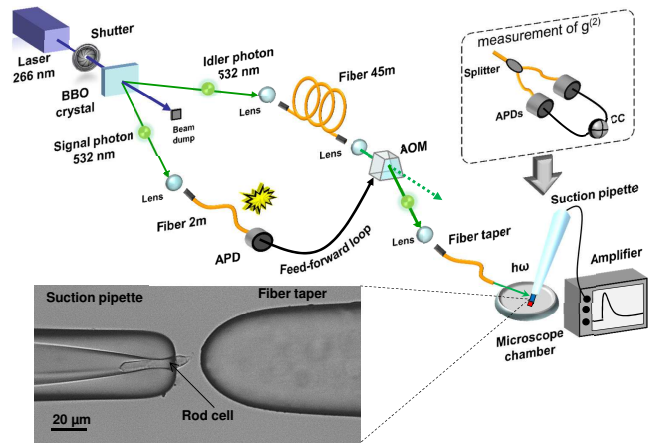}
			\caption{(Color online) Experimental setup. Photon pairs are produced via the spontaneous parametric down conversion in the $\beta$-barium borate (BBO) crystal pumped by a UV laser. The signal photon is detected by the avalanche photodiode (APD), and its output triggers an acousto-optical modulator (AOM). The idler photon is delayed by the fiber, and then diverted by the AOM to a fiber taper pointing at a rod cell. Electrical currents of the rod cell are measured by the technique of suction pipette. Single photons in the idler beam are characterized in a separate experiment by measuring $g^{(2)}$ using a 50/50 beamsplitter, two APDs, and a coincidence circuit (CC), see inset. Microscope (20X) image shows the rod cell in the suction pipette and the fiber taper in the recording configuration. Their positions are carefully aligned to ensure optimal light coupling.}
	\label{setup}
\end{figure}

Methods of cell preparation, electrophysiology recordings, and light coupling are similar to the ones we described previously \cite{SimBOEX, SimPRL}. Rod cells are obtained from dark-adapted adult male frogs (\textit{Xenopus laevis}) \cite{IACUC}. Rod cells  are loaded into a chamber of the inverted microscope placed in a light-tight Faraday cage. The microscope is equipped with IR lamp and a CCD camera. The membrane current of the rod cell is measured with the electrophysiological technique of suction pipette \cite{Baylor}. Pipettes are pulled from glass capillary and their tips have openings in a range between 6 $\mu$m to 7 $\mu$m. %Resistance of the pipettes in Ringer solution ranges from 1.5 $M\Omega$ to 2.0 $M\Omega$. It grows to 9 $M\Omega$ to 12.5 $M\Omega$ once the rod cell is sucked inside. 
The pipette is connected to the amplifier (Heka). Current waveforms are recorded with 100 Hz bandwidth, and along with trigger pulses from the APD, they are saved to the computer for subsequent analysis.

The rod cell is held in a glass pipette, and a taper of an optical fiber (Nanonics) is positioned next to it, see Fig.1. The light from the fiber propagates axially to the rod cell. \textcolor[rgb]{0,0,0}{Such arrangement allows to maximize photon absorption by the rod cell, and mimics the way light travels in the eye }\cite{SimBOEX}. The taper has a working distance of 22 $\mu$m, and a spot size of 4 $\mu$m, chosen to match the size of the cell. Selection of the responsive rod cells and control of their functionality is described in details in Supplemental Material \cite{Sup}. The experiments are conducted at room temperature ($20^{\circ}$C). Results obtained from ten rod cells from ten different animals are presented.

Initially a shutter blocks the pump beam, and the membrane current of the rod cell in the dark is recorded for 600 ms. The shutter is opened for 100 ms, and current is recorded for 5 s.  Waveform amplitude is calculated as a difference of the time-averaged membrane current at the peak of the response and at the baseline. Positions of time windows are defined individually for each rod cell by analyzing responses to pulses of an auxiliary laser. For each opening of the shutter the APD may or may not produce a photocount. Waveforms, accompanied by only a single APD photocount, are used to analyze single photon responses. Waveforms, accompanied by zero photocounts, are used to analyze the dark noise. Single photon responses and the dark noise are measured concurrently.

Probability distribution of waveform amplitudes for the case when the APD heralds a single photon, is shown in Fig.2A. It has asymmetrical shape with the mean 0.07 pA, and the variance 0.1 $pA^{2}$. A non-response peak, centered at 0 pA, corresponds to events when the rod cell fails to detect a photon or the photon was lost in the idler beam. A single photon response peak, centered at 0.58 pA, corresponds to successful single photon detection events. The histogram is fitted  by a sum of two Gaussian peaks centered at 0 pA and at 0.58 pA; both have a full width at the half maximum (FWHM) 0.5 pA. The fit yields coefficient of determination $R^{2}$=0.92. The peaks partially overlap due to the experimental noise, which includes contributions from continuous and discrete components of the physiological noise of the rod cell \cite{Baylornoise,Riekenoise}, and the Johnson noise in the seal resistance. 

Because of relatively small amplitudes of single photon responses for \textit{Xenopus} toad \cite{Xenopus1,Xenopus2}, it was not possible to clearly separate them from the experimental noise. More clear separation of the single photon peak would be possible with \textit{Bufo Marinus} toads \cite{Baylorsingle}. At the same time, use of controllable single photon stimulation guarantees that the observed asymmetry in the response histogram is caused by single photon detection, even in case if it is not clearly separated from the noise.

In Fig.2A multiphoton responses are not observed, and their statistics follows the statistics of light with a single photon precision \cite{SimPRL}. Note that in order to minimize the contribution of multiphoton responses with a classical light source, it would be necessary to adjust its light strength in accordance with the quantum efficiency of each cell. The latter  is not known in advance, and may vary significantly due to biological factors. Use of our light source allows to exclude  bias in assessing single photon responses for different cells, since it always provides strong suppression of the multiphoton component.

The distribution of dark noise amplitudes, shown in Fig.2B, has the mean 0 pA, and variance 0.07 $pA^2$. It shows convolution of the physiological noise of the rod cell with the noise of the recording system \cite{Baylornoise,Riekenoise}. The curve is fitted by a single Gaussian peak centered at 0 pA with FWHM=0.59 pA ($R^{2}$=0.97). 

A criterion-based method is used to identify single-photon responses. Waveforms with amplitudes higher than the criterion level are categorized as \textit{“single photon responses”} and lower than the criterion level as \textit{“non-responses”}. Based on the measurement of the noise of the amplifier, see Fig.S3 of Supplemental Material \cite{Sup}, the criterion level is set at 0.45 pA. 

We apply the amplitude threshold criterion ($>$0.45 pA) and sum all the probabilities for responses satisfying the criterion. The probability of occurrence of single photon responses is higher when the APD heralds a single photon, compared to the dark noise, see Fig.2C. The hypothesis is tested with Welch's unpaired t-test \cite{Welsh}. The one-tailed \textit{P} value is 0.028 for cell \#1, 0.00015 (\#2), 0.039 (\#3), 0.006 (\#4), 0.0001 (\#5), 0.053 (\#6), 0.0005 (\#7), 0.005 (\#8), 0.003 (\#9), 0.006 (\#10). Therefore, responsiveness of the cells to stimuli, produced by the single photon source is justified. \textcolor[rgb]{0,0,0}{Thus, we provide a model-independent proof of single photon sensitivity of rod cells, which was never attempted before.} The cell-to-cell variations are mainly attributed to intrinsic differences of cells to respond to single photons, because they originated from different animals and were obtained from different parts of the retina. 

\begin{figure} [h]
			\includegraphics[width=\linewidth]{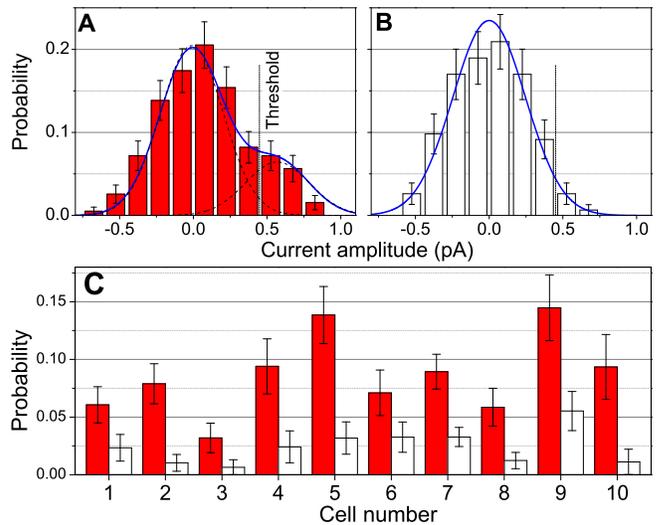}
						\caption{(Color online) (A) Probability distribution of amplitudes of rod cell responses when the APD in the signal beam heralds a single photon (n=195) and (B), for the dark noise (n=157). Solid lines are Gaussian fits. The vertical dash lines indicates the criterion level for categorization of single photon responses. (C) Overall probability of occurrence of single photon responses, satisfying the criterion, when the APD heralds a single photon (red bars), and for the dark noise (white bars). The total number of experimental trials is 402 for cell \#1, 435 (\#2), 342 (\#3), 273 (\#4), 352 (\#5), 353 (\#6), 816 (\#7), 449 (\#8), 333 (\#9), 197 (\#10). Error bars in (A-C) show $\pm$ s.d. Plots in (A, B) correspond to cell \#5 in (C).}
	\label{setup}
\end{figure}

Averaged waveform of single photon responses and non-responses for cell \#5 are shown in Figs.3A, B. It is fitted with the impulse response of the Poisson filter $i(t)=A_{0}[t/t_{0} exp⁡(1-t/t_{0})]^{(m-1)}$ with the amplitude $A_{0}$=0.58 pA, number of stages m=4, and time to peak $t_{0}$=1.75 s \cite{Baylorpulse}. Waveform parameters for all the studied cells are shown in Fig.S4 of Supplemental Material \cite{Sup}. The responses have the amplitude (0.59 $\pm$ 0.01) pA, time-to-peak (1.8 $\pm$ 0.2) s, and duration at the full width at half maximum (2.2 $\pm$ 0.2) s (mean $\pm$ s.e.m. n=10). The mean values are close to the ones observed in experiments with conventional light sources and rod cells from the same species \cite{Xenopus1,Xenopus2}. However, due to use of controllable single photon source, the dependence of the observed fluctuations of response parameters on the number of incident photons is excluded. \textcolor[rgb]{0,0,0}{This opens the opportunity to directly assess intrinsic noise of the phototransduction process in rod cells, which is hindered by the presence of multiphoton events in experiments with classical light sources.\cite{Teich, Teich1}.}
\begin{figure}  [h]
\includegraphics[width=\linewidth]{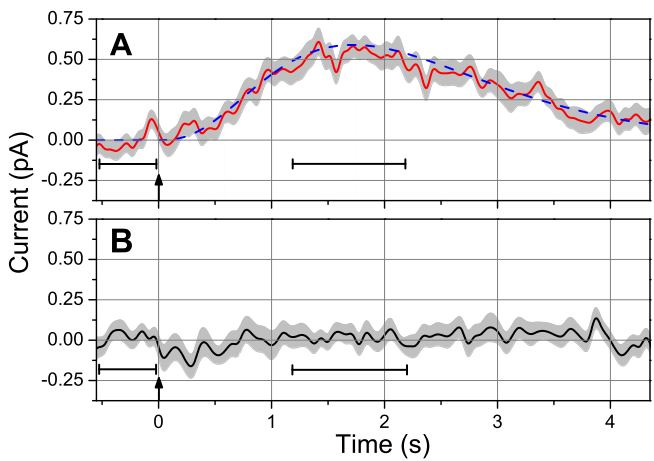}
\caption{(Color online) (A) Average waveform of the cell single photon responses (red solid line), and (B) of non-responses (bandwidth 20 Hz, n=27). Blue dashed line in (A) is the theoretical curve. Arrow indicates the moment of opening of the shutter in the pump beam. Horizontal bars show time windows for calculation of waveform amplitudes. Grey shaded regions in (A, B) show $\pm$ s.e.m. Plots in (A, B) correspond to cell \#5 in Fig.2C.}
\end{figure}

The quantum efficiencies \textcolor[rgb]{0,0,0}{of rod cells are} calculated from data in Fig.2C by taking into account optical losses in the idler channel and the rod cell dark noise, according to Eq.(S2) of Supplemental Material \cite{Sup}. The result yields $\eta$=(29 $\pm$ 4.7)\% (mean $\pm$ s.e.m. n=10). In earlier experiments  the rod cell of \textit{Bufo marinus} toad was illuminated by a transverse stripe of light \cite{Baylorsingle}. The probability of photon absorption was measured as 11.9\%, and the efficiency of the response to the absorbed photon was measured as 50\%, which result in $\eta$=6\%. Our results are consistently ($\approx$ 5 times) higher than this, since we use an axial geometry of photon delivery, and the proper single photon source, \textcolor[rgb]{0,0,0}{which excludes presence of multiphoton events}. Interestingly, that our result is close to the estimate of \cite{Review} for human rod cells, obtained from behavioral experiments. Note that in SPDC, wavelengths of signal and idler photons can be tuned in a very broad range \cite{KlyshkoPhotons}. Hence, it is possible to use this approach for measurement of spectral dependence of the quantum efficiency. \textcolor{black}{Moreover, since the results obtained by the method do not depend on the used equipment, its implementation would ensure integrity and credibility of relevant biological data emerging from different labs.}

In conclusion, we performed an experiment, where we send light pulses from a true single photon source, based on the SPDC, to the rod cell via an optical fiber, and measure the rod cell responses. We provide a direct \textcolor[rgb]{0,0,0}{and unambiguous} proof of single photon sensitivity of rod cells, characterize their single photon responses without any interference from multiphoton events, and measure their quantum efficiencies without using any pre-calibrated devices. Our approach is universal \textcolor[rgb]{0,0,0}{and direct }as it is not based on any particular statistical model of the cell response \textcolor[rgb]{0,0,0}{and it does not involve any indirect assumptions}. 

The approach can be directly extended to study responses of the whole visual system to controllable multiphoton stimulation, see Supplemental Material \cite{Sup} for the details. Such an experiment will allow precise determination of the visual threshold \cite{Teich,Holmes} and address a fundamental question about manifestation of quantum effects in neurobiology \cite{Sekatski,Pomarico}. 

The presented approach opens a way for exploiting quantum light in studies of other photo-induced processes, such as photosynthesis. The ultimate control over photon statistics could lead to new clues about manifestation of quantum coherence in such processes \cite{Engel,Collini,Brumer}. \textcolor{black}{From an engineering stand point it could help to define the properties required for a single photon detector, mimicking natural detection, with retinal rod cells forming the basis.}

We thank Nigel Sim, Alex Tok and Mike Jones for their help at various stages of the project. We thank Gleb Maslennikov, Vadim Volkov, John Dowling and Trevor Lamb for valuable comments. The work is done with financial support of A-STAR Joint Council Office grant No. 1231AEG025.

\newpage
\section{Supplemental Material}

\subsection{Synchronization of the experiment} The schematic of the setup is shown in Fig.S1. A pulse generator (repetition rate 25 kHz, Tektronix) with two synchronized outputs is used to drive the laser, and to gate the detection system (pulse width $\tau$=70 ns). The output of the avalanche photodiode (APD) in the signal arm (pulse width $\tau$=35 ns) is sent to the AND logic gate (Phillips Scientific, time window 120 ns), where it is logically multiplied with the gate pulse from the generator. Gating strongly supresses dark counts of the APD, providing the signal-to-noise ratio $\approx$60. The output of the AND gate (pulse width $\tau$=120 ns) is used to trigger the acousto-optical modulator (AOM) in the idler beam. The delay from the moment the APD produces a photocount, till the AOM is fully activated (transmission in the first diffraction order is maximized) is about 190 ns. It is compensated by a 45 m long single mode optical fiber (Thorlabs) in the idler beam, which introduces a delay of $\approx$230 ns.

\begin{suppfigure} [h]
			\includegraphics[width=\linewidth]{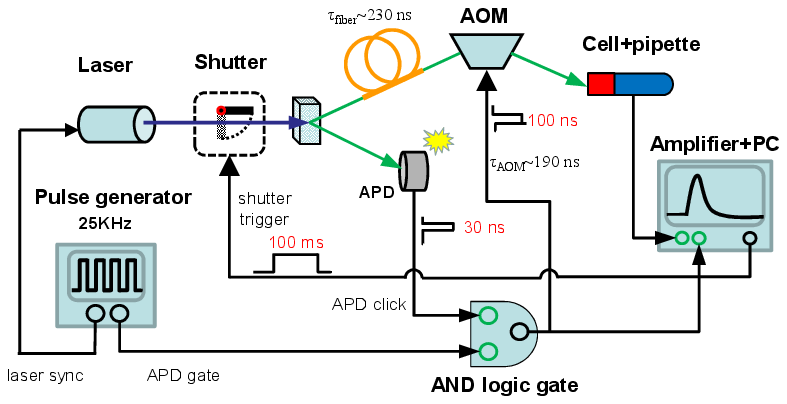}
			\caption{(Color online) Synchronization of the experiment. The pulse generator produces a sync pulse for the pump laser and the gate signal for the APD. The AND logic gate is used to gate the APD in order to reduce false photocounts. The output of the AND gate is fetched to the AOM, and the amplifier. The amplifier records the membrane current of the rod cell and the heralding signal from the APD. Once the acquisition is started, the amplifier produces a trigger signal for a mechanical shutter in the pump beam.}
	\label{sync}
\end{suppfigure}

Acquisition of current waveforms from the rod cell is started by the signal from the patch clamp amplifier. For the first 600 ms the dark noise of the rod cell is recorded. The amplifier produces a trigger pulse which opens the shutter in the pump beam for 100 ms. The current is recorded for 5 s, after which a new trigger pulse is sent to the shutter.

\subsection{Characterization of the single photon source} Normalized Glauber’s correlation functions (CF) are widely used for accessing photon statistics. Significant advantage of CF measurement over direct measurements of photon number distribution is that CFs are not sensitive to optical losses and quantum efficiencies of the detectors \cite{KlyshkoFoundations,Malte}. The light source is characterized by measuring the dependence of the second order CF $g^{(2)}$ on the pump power in three different configurations. 

\begin{enumerate}
	\item $g^{(2)}$ is measured between signal and idler beams with inactive feed forward loop. One places a single APD in each beam, and addresses their outputs to a coincidence circuit. $g^{(2)}$ is given by:
\begin{equation}
	g^{(2)}=f\times N_c/(N_1\times N_2) \tag{S1}
\end{equation}
where $N_1$, $N_2$ are the numbers of photocounts of APDs in signal and idler beams, $N_c$ is the number of photocounts coincidences, and $f$=25 kHz is the repetition rate of the laser. The corresponding dependence is shown in Fig.S2 by black symbols. High values of $g^{(2)}$ indicate strong pairwise correlations between signal and idler photons.
	\item $g^{(2)}$ is measured in the idler beam with inactive feed forward loop. One places a 50/50 fiber beam splitter in the idler beam with its outputs plugged to two APDs . The APD outputs are addressed to a coincidence circuit. $g^{(2)}$ is calculated according to Eq.(S1) but $N_1$, $N_2$ are the numbers of photocounts of two APDs in the idler beam. The corresponding dependence is shown in Fig.S2 by green symbols. The $g^{(2)}$ value is close to unity, and it does not depend on the pump power. This indicates that the photon statistics in the idler beam is Poissonian. This is expected since the SPDC source operates in a multimode regime.
	\item$g^{(2)}$ is measured in the idler beam with active feed forward loop (configuration realized in the experiment). The configuration is similar to the above case, but the pulses from the APD in the signal beam trigger the AOM in the idler beam. $g^{(2)}$ is calculated using Eq.(S1), but $f$ is the number of photocounts of the APD in the signal beam. The corresponding dependence is shown in Fig.S2 by red symbols. To ensure high fidelity of prepared single photon pulses, the experiments are conducted at the values of the pump power in the range from 4 $\mu$W to 7 $\mu$W. 
\end{enumerate}

\begin{suppfigure} [h]
			\includegraphics[width=\linewidth]{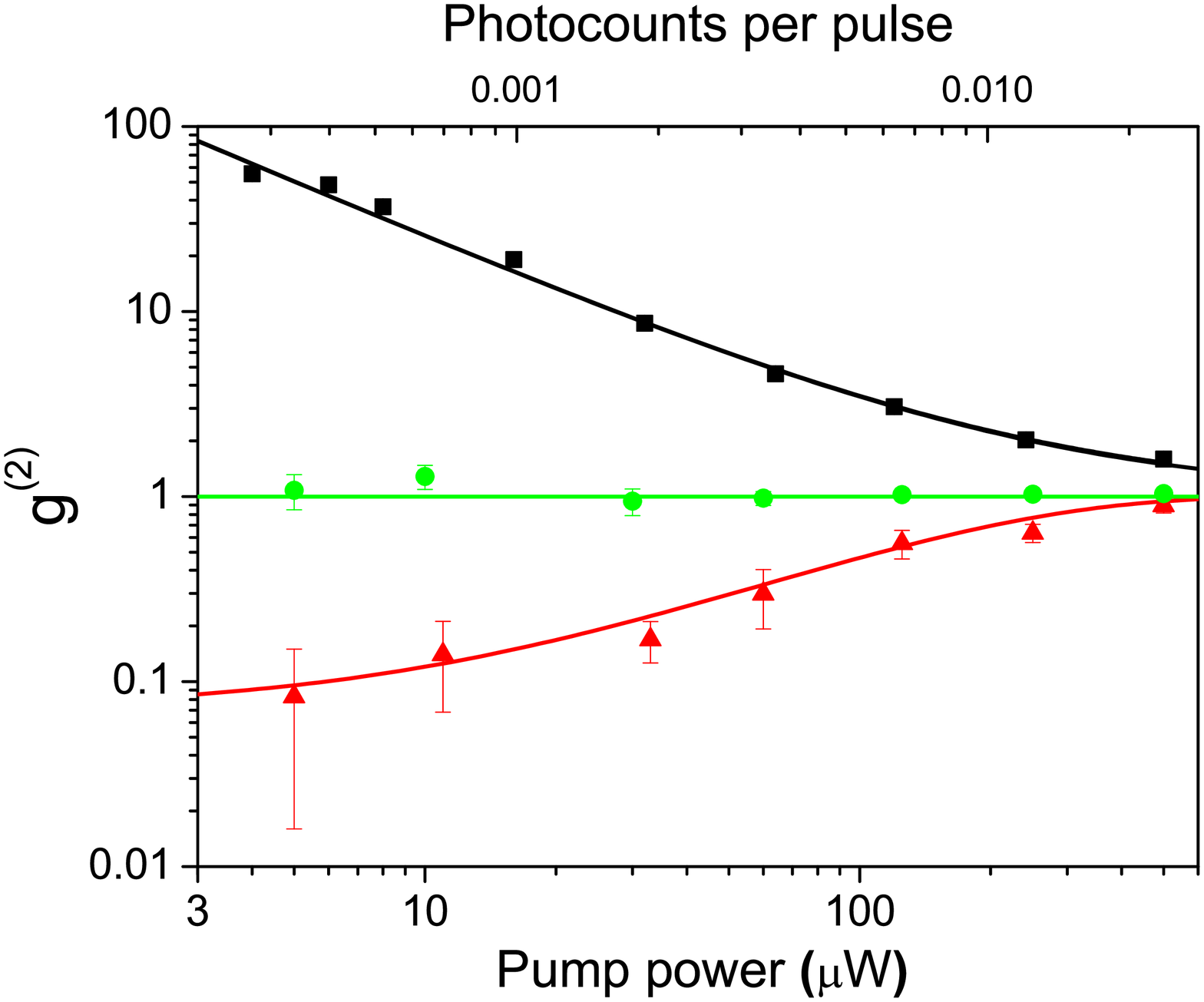}
			\caption{(Color online) Dependence of the second order intensity correlation function $g^{(2)}$ on the pump power,  measured: between signal and idler beams (black squares); in the idler beam, with inactive feed forward loop (green circles); in the idler beam, with active feed forward loop (red triangles). The axis on the top shows the corresponding number of photocounts of the APD in the signal beam per pulse of the pump laser. Lines are theoretical curves \cite{Malte}. Error bars are $\pm$ s.d.}
	\label{g2}
\end{suppfigure}

\subsection{Noise of the amplifier} The noise of the amplifier is measured using a 10 M$\Omega$ test circuit (Heka) attached to the input of the amplifier. The resistance of the circuit is chosen to mimic the pipette with the rod cell. The amplitude probability histogram is measured using same time windows, as those, used for the analysis of rod cell responses. The dependence is centered at 0 pA, and has a variance 0.032 pA$^2$, see Fig.S3. It is fitted by a Gaussian curve using Levenberg-Marquardt algorithm (Origin Lab). The fitting curve is centered at 0 pA, has FWHM=0.4 pA, and yields $R^2$=0.986. The criterion level for categorization of single photon responses is chosen at 0.45 pA. In this case the probability of observing the pulse with amplitude more than 0.45 pA is less than 1.1 \%. 

\begin{suppfigure} [h]
			\includegraphics[width=\linewidth]{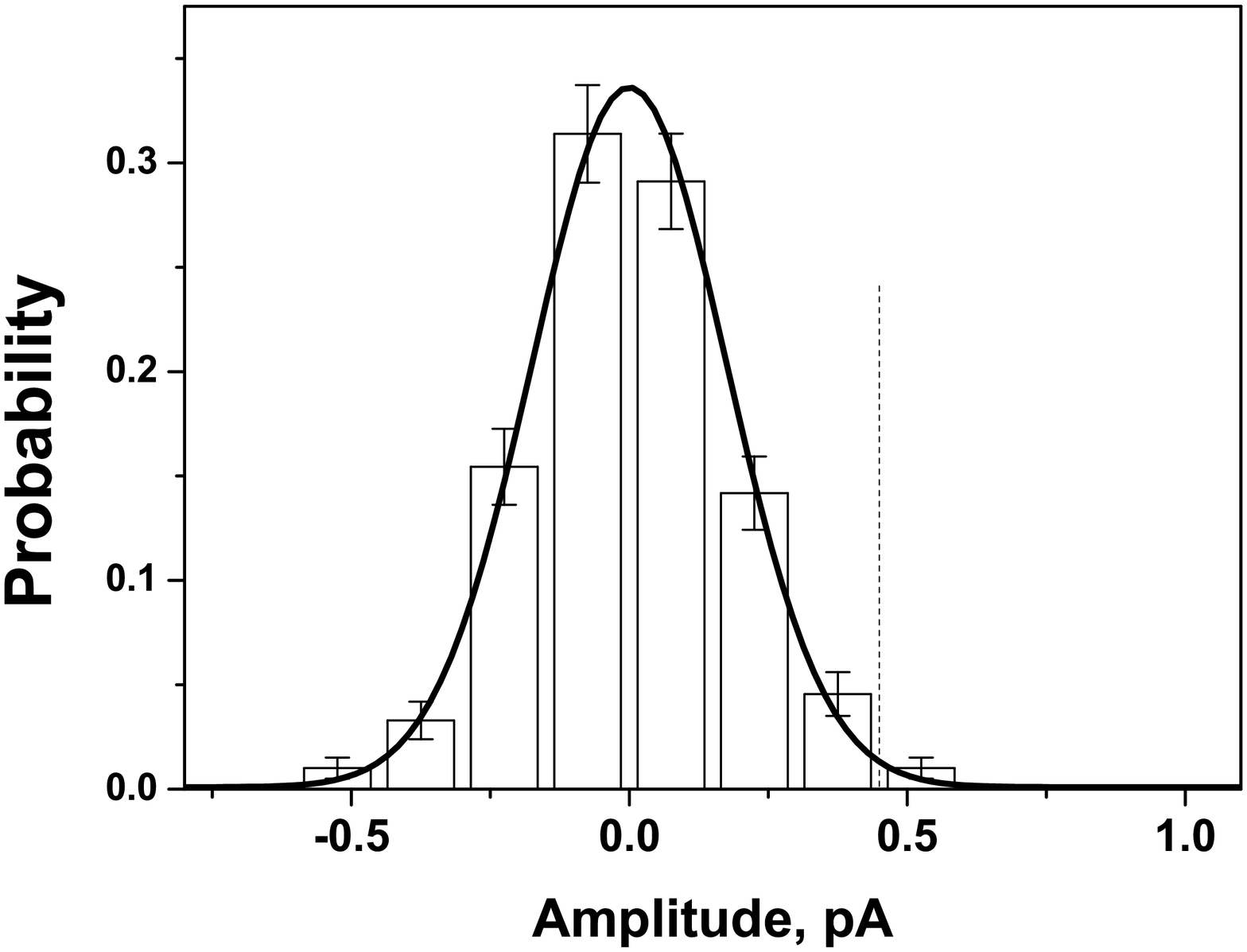} 
			\caption{Noise of the amplifier measured with the 10 M$\Omega$ test circuit (n=395). Bars are experimental data. Solid line is a Gaussian fit. Vertical dashed line shows the criterion level of 0.45 pA. Error bars show $\pm$ s.d. }
	\label{noise}
\end{suppfigure}

\subsection{Choice of functional rod cells} Since the experiments are conducted at the fundamental limit of light intensity (one photon at a time), selection of responsive rod cells and control of their functionality are crucial. Rod cell responses are probed by sending 5 ms pulses of an auxiliary 532 nm laser via the fiber taper. The laser intensity is adjusted to initiate rod cell response of half saturating amplitude (typically from 8 pA to 10 pA). Distinguishable single photon responses are observed for the rod cells, for which half-saturation response is initiated by no more than 100 to 250 photons per pulse. Functionality of the rod cell during the experiment is checked every 20 min by observing responses to the dim laser pulse of fixed intensity. Each rod cell is used for continuous recordings for 100 min to 120 min.

\subsection{Parameters of single photon responses} Measured parameters of single photon responses (amplitude, time-to-peak, and full width at the half amplitude) for 10 studied cells are shown in Fig.S4. The responses have the mean amplitude (0.59 $\pm$ 0.01) pA, time-to-peak (1.8 $\pm$ 0.2) s, and duration at the full width at half maximum (2.2 $\pm$ 0.2) s (mean $\pm$ s.e.m, n=10).

\begin{suppfigure} [h]
			\includegraphics[width=\linewidth]{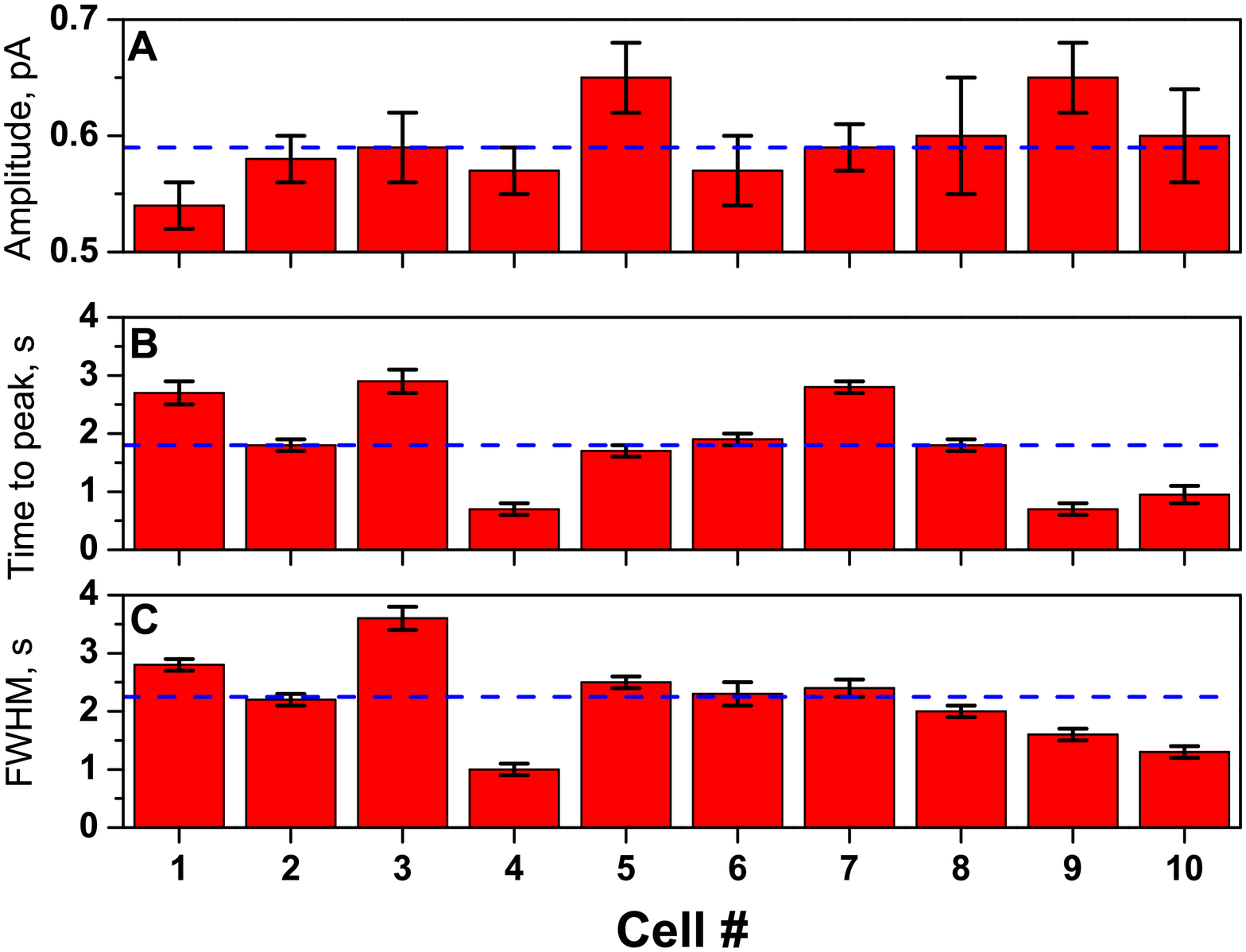}
			\caption{(Color online) Amplitude and temporal parameters of single photon responses for 10 investigated cells. (A) Mean amplitude; (B) Mean time-to-peak; (C) Mean duration at a full width at a half amplitude (FWHM). Error bars show $\pm$ s.e.m. Blue dashed lines show the median value (n=10).}
	\label{responses}
\end{suppfigure}

\subsection{Calculation of quantum efficiency} Intensity of the auxiliary 532 nm laser is measured at the input ($I_{in}$) and at the output ($I_{out}$) of optical elements in the idler beam using a power meter (Thorlabs). The transmission coefficient for each element is defined as $\eta=I_{out}/I_{in}$. The AOM efficiency is $\eta_{AOM}=60\%$, coupling efficiency into a single mode fiber taper is $\eta_{taper}=70 \%$. Propagation and coupling losses for the SPDC in the 45 meters long SM fiber were determined in the following way. First, the number of coincidences in the SPDC experiment was measured using the APD in the signal beam directly connected to the output of the SM fiber. Then the SM fiber was carefully replaced from the recipient coupler by a 2 m long multi-mode fiber (~50 $\mu$m core size), the APD was connected to the MM fiber, and the coincidences were measured again. The ratio of the two coincidences values gives $\eta_{fiber}=50 \%$. Quantum efficiency of the rod cell, corrected for the dark noise and optical losses in the idler beam, is calculated as follows:
\begin{equation} 
\eta=\frac{p_{sph}-p_{dn}}{\eta_{taper}\eta_{AOM}\eta_{fiber}}, \tag{S2}
\end{equation}
where $p_{sph}$, and $p_{dn}$ are the probabilities of occurrence of single photon responses when the APD heralds a single photon, and for the dark noise, respectively, see Fig. 2C.

\subsection{Multiphoton responses} During opening time of the shutter, 2500 pump laser pulses are “injected” into the setup. Because of probabilistic nature of the SPDC, several photon pairs, produced from different pump pulses, may stimulate the cell. Such events are unambiguously identified by observation of multiple photocounts of the APD in the signal beam, which operates in this case as a photon number resolving detector (dead time $\approx$ 35 ns). Multiphoton responses are observed in $\approx$ 7 \% of the measurements, and in this experiment they are excluded from the analysis. By taking the multiphoton events into account, it is possible to use our setup for further studies of cell responses to controllable multi-photon stimulation.


\begin{references}

\bibitem{Bahor} \textcolor[rgb]{0,0,0}{M. A. Taylor \textit{et al.,} Biological measurement beyond the quantum limit, Nature Photonics \textbf{7}, 229–233 (2013).}

\bibitem{Teich} M. C. Teich, P. R. Prucnal, G. Vannucci, M. E. Breton, W. J. McGill, Multiplication noise in the human visual system at threshold - The Role of Non-poissonian quantum fluctuations. Biol. Cybern. \textbf{44}, 157-165 (1982).

\bibitem{Sekatski} P. Sekatski, N. Brunner, C. Branciard, N. Gisin, C. Simon, Towards quantum experiments with human eyes as detectors based on cloning via stimulated emission. Phys. Rev. Lett. \textbf{103}, 113601 (2009).

\bibitem{Pomarico} \textcolor[rgb]{0,0,0}{E. Pomarico \textit{et al.} Experimental amplification of an entangled photon: what if the detection loophole is ignored? \textit{New J. Phys.} \textbf{13} 063031 (2011).}

\bibitem{Thaheld} \textcolor[rgb]{0,0,0}{F. H. Thaheld, Can we determine if the linear nature of quantum mechanics is violated by the perceptual process? \textit{BioSystems} \textbf{71}, 305-309 (2003).}

\bibitem{Koch}	Ch. Koch, K. Hepp, Quantum mechanics in the brain. Nature \textbf{440}, 611-612 (2006).

\bibitem{Baylor} D. A. Baylor, T. D. Lamb, K.-W. Yau, The membrane current of single rod outer segment. J. Physiol. (London) \textbf{288}, 589-611 (1979).

\bibitem{Bodoia} R. D. Bodoia, P. B. Detwiler, Patch-clamp recordings of the light-sensitive dark noise in retinal rods from the lizard and frog. J. Physiol. (London) \textbf{367}, 183-216 (1985).



\bibitem{Baylorsingle} D. A. Baylor, T. D. Lamb, K.-W. Yau, Responses of retinal rods to single photons. J. Physiol. (London) \textbf{288}, 613-634 (1979).

\bibitem{Review}	F. Rieke, D. A. Baylor, Single-photon detection by rod cells of the retina. Rev. Mod. Phys. \textbf{70}, 1027-1036 (1998).
%
%\bibitem{Hamer}	R. D. Hamer, S. C. Nicholas, D.Tranchina, T. D. Lamb, J. L. P. Jarvinen, Toward a unified model of vertebrate rod phototransduction. Visual neuroscience \textbf{22}, 417-436 (2005).
%
%\bibitem{Caruso} G. Caruso et al., Mathematical and computational modeling of spatio-temporal signalling in rod phototransduction. IEE Proceedings-Systems Biology \textbf{152}, 119-137 (2005).
\bibitem{Loudon} R. Loudon, \textit{The Quantum Theory of Light}. (Oxford University Press, New York, 2000).

\bibitem{Eisaman} M. D. Eisaman, J. Fan, A. Migdall, S. V Polyakov, Single-photon sources and detectors. Review of Scientific Instruments \textbf{82}, 071101 (2011).


\bibitem{Holmes}	R. Holmes, B. G. Christensen, W. Street, F. R. Wang, P. G. Kwiat, Determining the Lower Limit of Human Vision Using a Single-Photon Source, Quantum Electronics and Laser Science Conference 2012, QTu1E.

\bibitem{KlyshkoPhotons}	D. N. Klyshko, \textit{Photons and Nonlinear Optics} (Gordon and Breach, New York, 1988) pp. 285-327.

\bibitem{Harosi}	F. I. Harosi, Absorption spectra and linear dichroism of some amphibian photoreceptors. J. Gen. Physiol. \textbf{66}, 357-382 (1975).

\bibitem{Palacios} A. G. Palacios, R. Srivastava, T. H. Goldsmith, Spectral and polarization sensitivity of photocurrents of amphibian rods in the visible and ultraviolet. Visual Neurosci. \textbf{15}, 319-331 (1998).

\bibitem{Rarity}	J. G. Rarity, P. R. Tapster, E. Jakeman, Observation of sub-Poissonian light in Parametric Down Conversion. Opt. Commun. \textbf{62}, 201-206 (1987).

\bibitem{Sup}	See Supplemental Material for the details of Synchronization of the experiment, Characterization of the single photon source, Measurement of the noise of the amplifier, Choice of functional rod cells, Calculation of quantum efficiency, and Multiphoton responses   

\bibitem{KlyshkoFoundations}	D. N. Klyshko, \textit{Physical Foundations of Quantum Electronics} (World Scientific, Singapore, 2011) pp. 318-323.

\bibitem{Burnham} D. C. Burnham, D. L. Weinberg, Observation of simultaneity in parametric production of optical photon pairs. Phys. Rev. Lett. \textbf{25}, 84-87 (1970).

\bibitem{Malygin} A. A. Malygin, A. N. Penin, A. V. Sergienko, Absolute calibration of the sensitivity of photodetectors using a biphotonic field. JETP Letters \textbf{33}, 477-480 (1981).

\bibitem{Migdall} A. Migdall, Correlated Photon Metrology without absolute standards. Phys. Today \textbf{52}, 41-46 (1999).

\bibitem{SimBOEX}	N. Sim, D. Bessarab, C. M. Jones, L. A. Krivitsky, Method of targeted delivery of laser beam to isolated retinal rods by fiber optics. Biomed. Opt. Express \textbf{2}, 2926-2933 (2011).

\bibitem{SimPRL} N. Sim, M. F. Cheng, D. Bessarab, C. M. Jones, L. A. Krivitsky, Measurement of Photon Statistics with Live Photoreceptor Cells. Phys. Rev. Lett. \textbf{109}, 113601 (2012).

\bibitem{IACUC} All procedures with animals are carried under Institutional Animal Care and Use Committee (IACUC) regulations. 

\bibitem{Baylornoise}	D. A. Baylor, G. Matthews, K.-W. Yau, Two components of electrical dark noise in toad retinal rod outer segments. J. Physiol. (London) \textbf{309}, 591-621 (1980).

\bibitem{Riekenoise}	F. Rieke, D. A. Baylor, Origin and functional impact of dark noise in retinal cones. Neuron \textbf{26}, 181-186 (2000).


\bibitem{Xenopus1}	V. Kefalov, Y. Fu, N. Marsh-Armstrong, K.-W. Yau, Role of visual pigment properties in rod and cone phototransduction. Nature \textbf{425}, 526-531 (2003).

\bibitem{Xenopus2}	E. Solessio \textit{et al.}, Developmental regulation of calcium-dependent feedback in Xenopus rods. J. Gen. Physiol. \textbf{124}, 569-585 (2004).

\bibitem{Welsh}	B. L. Welch, The generalization of `Student's problem when several different population variances are involved. Biometrika \textbf{34}, 28-35 (1947).


\bibitem{Baylorpulse}	D. A. Baylor, A. L. Hodgkin, T. D. Lamb, The electrical response of turtle cones to flashes and steps of light, J. Physiol. (London) \textbf{242}, 685-727 (1974).

\bibitem{Teich1} \textcolor[rgb]{0,0,0}{M.C. Teich, et al., Multiplication noise in the human visual system at threshold: 1. Quantum fluctuations and minimum detectable energy, JOSA \textbf{72}, 419-431 (1982).}


%\bibitem{Burnham} D. C. Burnham, D. L. Weinberg, Observation of simultaneity in parametric production of optical photon pairs. Phys. Rev. Lett. \textbf{25}, 84-87 (1970).
%
%\bibitem{Malygin} A. A. Malygin, A. N. Penin, A. V. Sergienko, Absolute calibration of the sensitivity of photodetectors using a biphotonic field. JETP Letters \textbf{33}, 477-480 (1981).
%
%\bibitem{Migdall} A. Migdall, Correlated Photon Metrology without absolute standards. Phys. Today \textbf{52}, 41-46 (1999).

\bibitem{Engel} G. S. Engel \textit{et al.}, Evidence for wavelike energy transfer through quantum coherence in photosynthetic systems. Nature \textbf{446}, 783-786 (2007).

\bibitem{Collini} E. Collini \textit{et al.}, Coherently wired light-harvesting in photosynthetic marine algae at ambient temperature. Nature \textbf{463}, 644-648 (2010).

\bibitem{Brumer} P. Brumer, M. Shapiro, Molecular response in one-photon absorption via natural thermal light vs. pulsed laser excitation. Proc. Natl. Acad. Sci. U.S.A. \textbf{109}, 19575-19578 (2012).


\bibitem{Malte}	M. Avenhaus, K. Laiho, M.V. Chekhova, Ch. Silberhorn, Accessing Higher Order Correlations in Quantum Optical States by Time Multiplexing. Phys. Rev. Lett. \textbf{104}, 063602 (2010). 
 

\end{references}
\end{document}